# Adaptive Content Control for Communication amongst Cooperative Automated Vehicles

Mohammad Fanaei*, Amin Tahmasbi-Sarvestani*, Yaser P. Fallah*,
Gaurav Bansal†, Matthew C. Valenti*, and John B. Kenney†

* Department of Computer Science and Electrical Engineering, West Virginia University, Morgantown, WV, U.S.A.
† Toyota InfoTechnology Center, Mountain View, California, U.S.A.
Email: {mfanaei,amtahmasbisarvestani}@mix.wvu.edu, yaser.fallah@mail.wvu.edu, gbansal@us.toyota-itc.com,
valenti@ieee.org, and jkenney@us.toyota-itc.com

*Abstract*— Cooperative automated vehicles exchange information to assist each other in creating a more precise and extended view of their surroundings, with the aim of improving automated-driving decisions. This paper addresses the need for scalable communication among these vehicles. To this end, a general communication framework is proposed through which automated cars exchange information derived from multi-resolution maps created using their local sensing modalities. This method can extend the region visible to a car beyond the area directly sensed by its own sensors. An adaptive, probabilistic, distance-dependent strategy is proposed that controls the content of the messages exchanged among vehicles based on performance measures associated with the load on the communication channel.

## I. INTRODUCTION

An automated-driving system relies extensively on the vehicle's local sensors, such as radar, lidar, or cameras, to learn about its surroundings and make optimal driving decisions. The intrinsic limitations of these sensors in terms of range and field of view often prevent planning and maneuvers that would require information from farther distances and/or occluded spaces. Short-range wireless communication among vehicles could be used as a method to mitigate this issue and expand the reach of the sensory information locally available at each vehicle. Each such cooperative automated car can use all its sensing modalities to create a map containing information such as the location, speed, and heading associated with the cars, motorcycles, pedestrians, and other objects in its immediate neighborhood. It can then communicate (potentially a condensed version of) this map to other cars that cannot directly observe all of the same information, in order to expand their maps. Each vehicle will then be able to augment its map with the information it has received from other cars. Ultimately, the performance of any higher-level application using this map data, such as long-term trajectory planning, crash avoidance strategy, and eventually automated and autonomous driving, could be drastically improved through this cooperative exchange of local information.

Dedicated short-range communication (DSRC) is a key technology to enable communication between vehicles in emerging networked vehicular applications [1]. It has been shown that DSRC throughput degrades at high channel loads, e.g., due to an increase in the vehicle density, transmission rate or range, or the amount of information to be exchanged (i.e., packet size). In the scenario outlined above, it can be shown that even for a moderate vehicle density, the communication requirements of a network of automated cars cannot be satisfied if each car communicates its complete map frequently. Therefore, it is apparent that some discretion should be used in sharing information. This paper proposes a framework for a context-aware system that adapts rate, range, and/or content of the shared information based on the observed channel state. We also propose an adaptive, probabilistic, distance-dependent content-control strategy that can satisfy the requirements of our proposed framework.

Emerging cooperative vehicle-safety systems that rely on the broadcast of information (such as GPS data, path history, and path prediction) over DSRC networks are known to suffer from scalability challenges [2], even when each vehicle broadcasts the information related only to its own movements and not that of any other vehicles or objects. There are several solutions that are currently under study by the industry to address these scalability issues [3][4][5]. Given that automated vehicles rely heavily on rich sensory information, sharing large amounts of map data poses an even greater challenge. Therefore, a new communication framework and novel application-specific mechanisms are needed to resolve this problem in practical vehicular networks.

Recently, a communication system has been proposed in [6] to address the requirements of automated driving by allowing vehicles to request a multi-resolution map representation of any region in their surrounding environment. The proposed content-centric MAC scheme provides a mechanism through which the limited wireless bandwidth is adaptively allocated to the information objects based on the number of requests for them and to the vehicles based on the timeliness and completeness of the information that they have about the requested region.

We propose the separation of higher-layer decision-making applications, such as the automated and autonomous driving applications, from the sensing and communication modules through *real-time situational-awareness maps*. Current approaches proposed in the literature for the formation of these maps use only the *local* sensory data collected at each vehicle (e.g., [6]). This paper focuses on scalable communication solutions for exchanging such information among vehicles and updating local maps based on the fusion of locally sensed information as well as the data received from other cars. Our objective is to devise a scalable communication framework for this application and to propose practical control algorithms to improve the performance and scalability of the inter-vehicle communications.

The rest of this paper is organized as follows: Section II describes the system model of an automated vehicle and shows the extent of the scalability problem in communications among cars. Our proposed communication framework to resolve this scalability problem is introduced in Section III. Section IV presents details of a probabilistic, distance-dependent content-control strategy to adaptively adjust the map information in the messages broadcasted by each vehicle, based on the conditions of the communication medium. Section V presents results from NS-3 simulation experiments and evaluations of the proposed content-control scheme. Finally, the paper is concluded in Section VI.

## II. SYSTEM MODEL AND PROBLEM STATEMENT

A general model for an automated vehicle is shown in Figure 1, emphasizing on the role of situational awareness. Each car is equipped with a set of local sensing modalities such as lidar, radar, cameras, and GPS. The information collected by these sensors can be used to create a real-time, three-dimensional map of the surrounding environment of the car up to the sensing distance of local sensors. This map can effectively be used in applications such as lane-keep assist system, advanced pre-collision alert system, adaptive cruise control system, or automated crash-avoidance system. Note that these systems depend *only* on the information about the immediate neighborhood of a car. However, other higher-layer applications, such as long-term trajectory planning and automated and autonomous driving applications, require the above map to expand beyond the sensing distance of local sensors. For example, an automated car needs to have information about the traffic at the intersection that it will pass within the next ten seconds. This type of information is not readily available from the local sensors deployed on intelligent cars today.

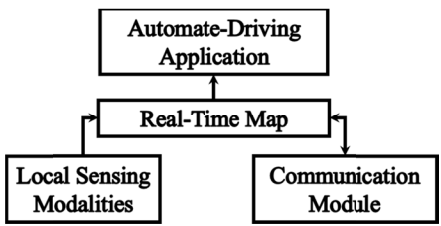

Figure 1: System model of a cooperative automated vehicle

The solution for the above problem is to extend the map locally available at each vehicle by exchanging map information among cars close to each other. Each car could broadcast a condensed version of its own local map containing information that either is explicitly requested by another car or is timely, accurate, and likely to be useful to other cars for their planning (e.g., the information about an unknown object in the middle of the road). Each vehicle could then combine its own observations through local sensor collections and the information received from other cars within its communication range to form and update its local map. The information extracted from the updated map can then be sent to the communication module in order to be disseminated to other neighboring vehicles. This scheme could expand the coverage of the maps so that the automated-driving application using the maps can perform more effectively.

As the size of the map information to be exchanged among vehicles is usually large, the above proposed scheme is not yet scalable and could easily lead to congestion in a network of DSRC transceivers. To highlight the problem of inter-vehicle communication scalability for a typical vehicular network, consider the following simulation environment. Suppose that vehicles move in a 4-Km long highway as studied in [7] and depicted in Figure 2. There are three lanes in each direction and a winding section in the middle in which the radius of curvature is 40 meters. The average speeds of the cars within the three lanes in each direction are 17 m/s, 18 m/s, and 19 m/s from the slowest (rightmost) to the fastest (leftmost) lane, respectively. A realistic mobility trace for vehicles is created in the traffic simulator SUMO with the modifications detailed in [7] for the SUMO's lane-change model and car-following model.

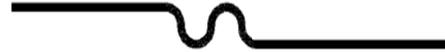

Figure 2: Road topology used in our NS-3 simulations [7]

In our NS-3 simulations, it is assumed that the wireless communication channels in the 5.9 GHz band experience two-ray path loss [8] (with the path-loss exponent of $\alpha = 2$ and parameter $\epsilon_r = 1.025$) and the default built-in NS-3 Nakagami fading (with distance-dependent parameter $m$). The MAC protocol is CSMA/CA with enhanced distributed channel access (EDCA). Other simulation parameters are summarized in TABLE 1. The transmission power is varied as 10, 21, and 31 dBm, which are equivalent to a transmission range of 250, 500, and 1000 meters, respectively. The transmission range is defined as the distance at which at least 90% of packets are successfully received, ignoring the two-ray null effect. In our simulations, cars do not include information about non-vehicular objects. Suppose that each car uses 260 bytes to include information about its own location, path history, and path predication (as in DSRC's basic safety messages [1]) and 60 bytes to represent the information regarding any other cars in its map. Therefore, the size of the message generated by each car would be $260 + 60x$ bytes, where $x$ is the number of cars within the transmitter's map. Assume that the car density in the road is 125 vehicles per kilometer and that each car only includes the information about other vehicles within its transmission range. The message size can then be variable as 260 (when no information about any other cars is included), 3980 (for 62 cars within the transmission range of 250 meters), 7760 (for 125 cars within the transmission range of 500 meters), and 15260 (for 250 cars within the transmission range of 1000 meters) bytes. When the message size exceeds the maximum packet length in the DSRC standard, the message is fragmented across multiple packets.

Figure 3 and Figure 4 show the *channel busy ratio* (CBR) versus the average offered load for different values of the transmission range, when the car density is 25 and 125 vehicles/Km on the 4-Km road topology shown in Figure 2, respectively. CBR values are calculated based on the approach described in [9] and averaged over the simulation time of 100 seconds and across all vehicles in the network. The average offered load is defined as $\Delta = f \times L$, where $f$ is the message transmission frequency of each vehicle and $L$ is the message size.

TABLE 1: SYSTEM PARAMETERS USED IN OUR NS-3 SIMULATIONS

| Parameter | Value | Parameter | Value |
|---|---|---|---|
| Noise Floor | -99 dBm | AIFSN | 2 |
| Carrier-sense Threshold | -94 dBm | Contention Window (min) | 15 |
| Packet Reception SINR | 7 dBm | Transmission Frequency ($f$) | 5 Hz (Fixed) |
| Channel Bandwidth | | | 10 MHz |

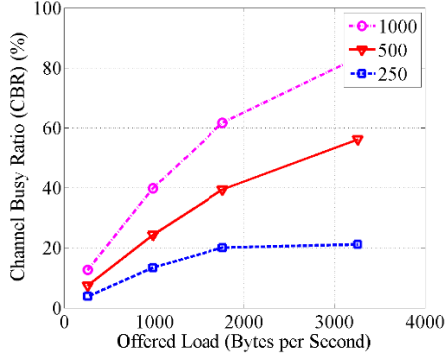

Figure 3: CBR versus average offered load for different values of the transmission range, when the car density is 25 vehicles per kilometer.

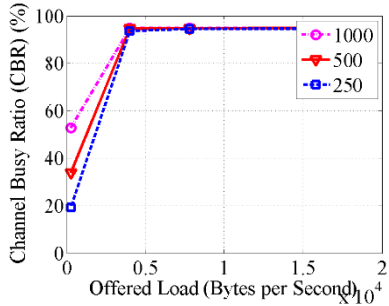

Figure 4: CBR versus average offered load for different values of the transmission range, when the car density is 125 vehicles per kilometer.

As shown in Figures 3 and 4, the CBR increases as the offered load or transmission range increases. For the case of 125 vehicles/Km, the CBR is unacceptably high even for a moderate offered load and transmission range. Note that a study in [9] found the acceptable range of the CBR values to be between 0.4 and 0.8 (ideally around 0.68). With a CBR above 0.8, there would be too many collisions, and a CBR below 0.4 indicates significant underutilization of the channel. When the CBR is outside of this desirable range, the *information dissemination rate* (IDR), which denotes the number of copies of a message that is successfully delivered, is drastically reduced [9]. In other words, when the CBR is outside of this range, the delivery rate for the map-update messages is significantly lower than what is ideally possible. Therefore, current communication strategies fail to satisfy the communication requirements of an automated-driving system. In the following sections, a scalable communication framework will be proposed that can adapt to the communication channel and implement a mechanism for exchanging map information among vehicles in a reliable manner.

## III. PROPOSED COMMUNICATION ARCHITECTURE

Cooperative automated vehicles rely on communications to extend their information about regions not directly sensed by their local sensing modalities. We propose to achieve this objective by separating the automated-driving applications and communication and sensing modules through real-time situational-awareness maps that are already used in automated cars. These maps are currently created and updated using only local sensors. We propose to enhance the maps using information from other cars. The map also feeds the communication module with information that can be disseminated to other cars. This model follows the current model of operation for cooperative vehicle-safety systems [2]; however, the information that is exchanged among vehicles is derived from a map and is much richer than simple GPS coordinates and internal sensors that are used in [2].

Figure 5 depicts a functional diagram of the proposed communication framework. Each car combines the information received from three different sources to create and update a multi-resolution local map used in any situational-awareness application. These sources are the information derived from local sensing modalities at each vehicle, previous knowledge about the driving environment available at each car, and the information derived through processing map data received from other vehicles in the network. Previous map knowledge can be in the form of the map information about permanent structures, buildings, roads, lanes, intersections, and traffic signs that can be pre-programmed in each vehicle or derived from previous versions of the map. The map at each car can be represented in a multi-resolution format such as the ones proposed in [6][10]. The resolution of the map depends on the required accuracy for the higher-layer applications using the map, available information from the surrounding environment, the reliability of the available information, and the storage and processing capabilities at each vehicle.

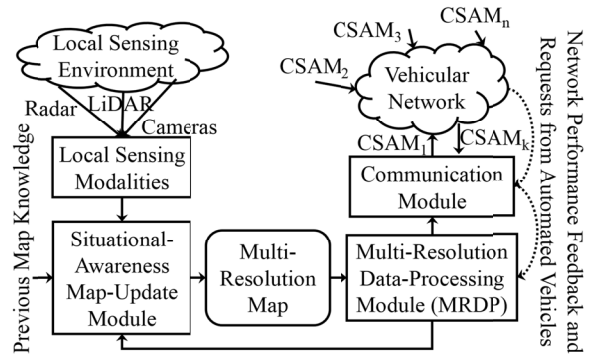

Figure 5: Proposed scalable communication architecture

The "multi-resolution data processing module" (MRDP) is an interface between the created map at each vehicle and the communication module. The MRDP module has two main functionalities: generating *cooperative situational-awareness messages* (CSAM) and extracting information from the received CSAMs from other vehicles. After the local multi-resolution map is updated, the MRDP module creates a message to be broadcasted in the communication channel based on the map content, feedback from the network performance, and any explicit requests received from other automated vehicles. In generating CSAM messages, the MRDP module considers the reliability of the information available in the local map, the current congestion and load on the communication network, and the content of the other CSAM messages and requests received from other cars. This architecture creates a distributed framework that can adapt to the conditions of the communication medium and resolve the scalability problem by avoiding repetitive transmissions of the same information by several cars. The MRDP module also processes the received CSAM messages from other vehicles, extracts reliable information from them, and sends the extracted information to the "map-update module" so that the resolution and/or range of the multi-resolution map created by this module can be enhanced.

There are different ways that the MRDP module can control the creation of CSAM messages based on the conditions of the communication channel. We categorize these methods into three different classes, namely *rate-control schemes*, *range-control schemes*, and *content-control schemes*. In a rate-control strategy, the transmission power is fixed and the rate at which updated map information is broadcasted adapts to the network conditions based on a measure such as the packet error rate (PER), channel busy ratio (CBR), or information dissemination rate (IDR). In a range-control scheme, the rate of broadcasting map updates is fixed and the transmission power is adapted in response to the aforementioned measures in order to improve network or system performance. For example, as the CBR increases beyond an acceptable threshold, the transmission power is decreased so that the interference created at larger distances and the load on the communication medium are reduced. In a content-control method, the content of messages exchanged among vehicles is adapted to the channel conditions. As the channel becomes less congested, each car can broadcast a higher-resolution version of its local map. This could affect both temporal and spatial accuracy of the maps created at different vehicles.

We believe that a scalable control strategy for automated-driving communications should ideally be a combination of *joint* message size, rate, and power (or range) control such that all three parameters are dynamically adapted to control the channel load. The message rate- and power-control schemes are extensively studied in the literature for the DSRC safety communications [2]-[11]. The focus of our work in this paper is to propose an adaptive message size- and content-control scheme for automated-driving communications. Presently and in order for a detailed analysis of content control, it is assumed that the transmission rate and power are fixed. As future work, we will investigate how message size can be jointly adapted with transmission rate and/or power.

## IV. Proposed Content-Control Scheme

The main idea of the proposed strategy is to control the *size* and *content* of the messages containing the map information of any vehicle based on the load on the network. Vehicles periodically broadcast their multi-resolution map information with a fixed rate and power. The size of the messages transmitted by an arbitrary vehicle is chosen so that its measured load of the network remains within a reasonable range. Having determined the message size, the content of the message is chosen in a probabilistic manner in such a way that the information regarding objects closer to the sender are included in the message with a higher probability. This method of map forwarding can be dubbed a "distance-sensitive" forwarding mechanism.

In order to formally introduce the scheme summarized above, the rest of this section is organized as follows: First, the *message size-control strategy* is described in detail to formulate how an optimal message size can be found for any given channel load. Then, a novel method is discussed to represent the content of a map available at an arbitrary vehicle. This map-representation scheme categorizes the objects in the map into known and unknown objects and represents them in an efficient, context-aware format. Finally, a probabilistic selection strategy is proposed to determine what kind of information about the objects in a map should be included in a message with the optimal size found in the first step.

### A. Message-Size Control Strategy

In our discussions, the ultimate goal is to deliver as much information as possible about a local map to as many vehicles as possible around the car possessing the map information. One performance measure that can capture both the amount of information successfully delivered and the number of receivers is the *information dissemination rate* (IDR), defined as "the number of copies of a node's packets that are successfully delivered" [9]. It has been shown in [9] that the IDR is a function of the locally measurable CBR. The results of [9] show that the IDR is maximized when the CBR is at 0.68 for a given linear highway topology, and that it stays in a reasonable range for the values of CBR in the interval of 0.4 to 0.8.

Based on the above discussion, we propose to have a message size-control strategy that adaptively adjusts the message size in such a way that the observed CBR by each vehicle remains near a desired value of $CBR^* = 0.68$. To this end, a descent-based message size-control algorithm is used that iteratively updates the optimal message size at iteration $t + 1$ as [12]

$$L_{OPT}^{t+1} = L_{OPT}^t + \eta \times (CBR^* - CBR^t)$$

where $L_{OPT}^t$ denotes the message size at iteration $t$, $CBR^t$ is the resulting observed channel occupancy, and $\eta$ is a tunable gain that can be adjusted to ensure convergence of the algorithm [12]. The above strategy ensures that each vehicle will adapt its message size such that the observed CBR in the network is near the desired value. This control mechanism reduces the message size as the CBR increases in order to prevent the CBR from exceeding the desired range. Conversely, when the CBR is lower than its desired value, the control mechanism increases the message size so that more information about the objects in the transmitter's map can be disseminated to other vehicles in the network.

The optimal message size should be bounded between two limits, namely $L_{min}$ and $L_{max}$. The minimum message size required for representing an arbitrary object in the map available at a vehicle is denoted by $L_{min}$. In other words, if a vehicle decides to transmit a message, it should contain the information about at least one object in the map. On the other hand, the maximum message size in our application is upper bounded by $L_{max}$, even if the CBR is low enough to support higher message sizes (the extra space could be used to increase the transmission frequency or range). Let $R$ and $f$ denote the physical-layer transmission rate (in bits per second) and the frequency of transmissions (in Hertz) at a vehicle. If the transmission overhead is denoted by $\gamma$ and the minimum number of transmitters in a close proximity is represented by $Q_{\min}$, the maximum message size at each transmitter (in bits) can be found as

$$L_{\max} = \frac{R}{fQ_{\min}}(1-\gamma)$$

where $Q_{\min}$ is determined based on the vehicle density in the driving environment within the communication range. As an example, if $R = 6$ Mbps, $f = 5$ Hz, $\gamma = 0.1$, and $Q_{\min} = 25$, the maximum message size is 43.2 Kbits, or 5,400 bytes. Note that the value of $Q_{\min}$ is chosen based on the assumption that the car density is 25 vehicles/Km and that the interferers are considered in a range twice as long as the transmission range of 250 meters.

*B. Representation of Map Information*

The objects in a given map are divided into two main categories: *known* objects and *unknown* objects. Let $K$ and $U$ denote the number of known and unknown objects in the map available at a typical vehicle, respectively. The information broadcasted for the $k$th *known* object is represented as

$$(\text{Type}_k, \Delta x_k, \Delta y_k, x_{C,k}, y_{C,k}, v_k, H_k, \theta_k, [\Omega_k]), k = 1,2,\dots,K,$$

where $\text{Type}_k$ is the type of the known object from a finite set of object types that a higher-level application can recognize from the vehicle's sensory data (e.g., car, truck, motorcycle, bicycle, and pedestrian), $\Delta x_k$ and $\Delta y_k$ are the object's $x$ and $y$ dimensions, respectively, $(x_{C,k}, y_{C,k})$ is the location of the center of the $k$th object, $v_k$ is the object's velocity, $H_k$ is the movement heading of the object, and $\theta_k$ is the orientation of the object around the $y$ direction (known as the *yaw* angle). In this representation, $\Omega_k$ denotes an *optional* set with a *fixed* length $M$, containing the path history and/or path prediction information about the $k$th known object. Each entry in $\Omega_k$ is in the form of $(x_{C,k,m}, y_{C,k,m}, v_{k,m}, H_{k,m}, \theta_{k,m}), m = 1,2,\dots,M$ containing the information about the $k$th object at the $m$th previous and/or future time slot.

Let $\ell_K$ denote the size of the information regarding any known object without considering the history/prediction data (i.e., everything except $\Omega_k$). Furthermore, suppose that $\ell_H$ is the size of the path history/prediction information $\Omega_k$ associated with a known object. If a message contains the reported information about $K_R$ known objects and the optional history/prediction information about a subset of them with $K_{R,h} \leq K_R$ objects, the size of the message can be found as $K_R \ell_K + K_{R,h} \ell_H$.

In order to represent the information associated with any *unknown* object in a map, an arbitrary multi-resolution approach can be used (e.g., [6][10]). For example, the following approach is proposed in this paper: The volume occupied by the $u$th unknown object is divided into $N_u$ cubic sub-regions, where the $n$th *occupied* sub-region is represented as

$$(x_{u,n}, y_{u,n}, z_{u,n}, D_{u,n}), n = 1,2,\dots,N_u \text{ and } u = 1,2,\dots,U,$$

where $(x_{u,n}, y_{u,n}, z_{u,n})$ is the center of the $n$th cube associated with the $u$th unknown object and $D_{u,n}$ is its dimension. As the resolution of the information associated with an unknown object increases, the number of sub-regions $N_u$ is increased and the dimension of each cubic sub-region $D_{u,n}$ is decreased.

Let $\ell_U$ denote the size of the information regarding each cubic sub-region of any unknown object. The size of the message containing reported information about $U_R$ unknown objects in the map of a typical vehicle can be found as $\ell_U \sum_{u=1}^{U_R} N_u$. If the number of the cubic sub-regions for all of the unknown objects is the same (i.e., $N_u \equiv N$), the size of the message containing reported information about $U_R$ unknown objects becomes $NU_R \ell_U$.

*C. Distance-Dependent Content-Control Strategy*

Suppose that a vehicle has measured the local CBR and determined the optimal message size for its map-update CSAM messages based on the strategy explained in Subsection IV.A. Furthermore, assume that it has formed its map and represented the objects in the map using the format described in Subsection IV.B. The next question would be "what is the strategy by which the vehicle chooses the content of its messages with the specified size?" The *probabilistic, distance-dependent content-control strategy* proposed in this subsection answers this question. In short, the proposed scheme is implemented through two main functionalities: First, the number of known and unknown objects whose information can be included in the message is determined. Then, the objects themselves are selected from the set of entire objects in the map in a probabilistic manner as described in the following.

Figure 6 summarizes the proposed strategy through which the number of objects whose information can be included in the message is determined. Let $K_R$ and $U_R$ be the number of known and unknown objects whose information is reported in the message, respectively. Furthermore, suppose that $K_{R,h} \leq K_R$ denotes the number of known objects for which the path history/prediction information is also included in the message. Note that the path history/prediction information of a known object is added to the message *only* if the main information associated with that object is already included. Based on the proposed scheme, the algorithm prioritizes the inclusion of information about as many unknown objects as possible at the lowest possible resolution. In other words, the algorithm starts by assigning only one cubic sub-region to each unknown object, i.e., $N_u = N \equiv 1, u = 1,2,\dots,U_R$. The next step is to include information about as many known objects in the map as possible without considering any of their history/prediction information. The remaining free space left after including the above information is used to add data about the history and/or prediction of movement for as many included known objects as possible. If there exists any further remaining space in the message, the resolution of the information associated with the unknown objects is increased by incrementing the number of cubic sub-regions assigned to each one of them and repeating the above process.

Having determined how many known and unknown objects could be included in a message based on the aforementioned strategy, the next step is to select that many objects from the set of all possibilities if there is space to include information for only a subset of objects. As discussed in Section III, besides the

```
Require: L_OPT, K, U, ℓ_K, ℓ_H, and ℓ_U.
1.  Initialization
2.    K_R ← 0, K_{R,h} ← 0, and U_R ← 0
3.    N ← 1 and N_OPT ← 0
4.    L_OPT^INIT ← L_OPT
5.  EndInitialization
6.  while L_OPT^INIT ≥ K_R ℓ_R + K_{R,h} ℓ_H + N U_R ℓ_U do
7.    L_OPT ← L_OPT^INIT
8.    U_R ← min(U, ⌊L_OPT/(Nℓ_U)⌋)
9.    L_OPT ← L_OPT − N U_R ℓ_U
10.   K_R ← min(K, ⌊L_OPT/ℓ_K⌋)
11.   L_OPT ← L_OPT − K_R ℓ_K
12.   K_{R,h} ← min(K_R, ⌊L_OPT/ℓ_H⌋)
13.   N_OPT ← N and N ← N + 1
14. end while
15. return K_R, K_{R,h}, U_R, and N_OPT.
```

Figure 6: Pseudocode for determining the number of objects whose information can be included in CSAM messages, where ⌊.⌋ denotes the floor operation.

locally observed network performance measure, the messages received from other vehicles in the network also affect the content of the messages that a vehicle broadcasts. In order to reduce the amount of *redundancy* in the proposed message-exchange protocol, each car first removes the *known* objects that it overhears their information broadcasted by other vehicles during the preceding time period $T$ from the pool of objects whose information can be included in the message. It also removes any *unknown* objects whose information has been broadcasted by other cars with a minimum resolution $N_{\min}$. If the number of remaining objects after this removal process is greater than the number of objects whose information can be included in the message, the following *probabilistic, distance-dependent* scheme is used to select a subset of them for transmission. Based on the proposed object-selection strategy, the message includes the information about objects whose distance to the transmitter is less than $r_0$ with probability 1. Beyond distance $r_0$, as the distance between the transmitter and the object in its map increases, the probability that the object's information is included in the broadcasted messages of the transmitter decreases monotonically. Therefore, it is more probable that a message contains information about closer objects than about a farther one. In this paper, the probability that an object located at distance $r$ from the transmitter is included in its broadcasted messages is assumed to be

$$\Pr(\text{Object Reported}) = \begin{cases} 1, & \text{if } r \leq r_0 \\ \lambda \exp(-\lambda r), & \text{if } r > r_0 \end{cases}$$

where $\lambda = \frac{1}{r_0} \ln \frac{1}{|1-r_0|}$.

Note that as the distance between a vehicle and an object decreases, the reliability of the available information about that object tends to increase, and the information about that object becomes more instrumental to the vehicle-safety and automated-driving applications. Hence, the proposed probabilistic approach implicitly emphasizes on the dissemination of more reliable and more important information.

## V. NUMERICAL EVALUATIONS

In this section, the results of NS-3 simulations are presented to show the scalability of the proposed algorithm. The simulation setup is similar to the one described in Section II. The parameters of the simulations have the same values as those outlined in TABLE 1. The vehicles are uniformly distributed in the 4-Km highway described in Section II. In order to concentrate on the effect of the proposed content-control scheme on the performance of the system, it is assumed that the frequency of the map updates is fixed at $f = 5$ Hz. For simplicity, all objects in the maps available at vehicles are assumed to be of known type, i.e., $U = 0$. The size of a message associated with the information regarding any known object without considering the history/prediction data is assumed to be $\ell_K = 60$ bytes (one integer field $\text{Type}_k$ and seven double fields). The size of the path history/prediction information $\Omega_k$ for a known object is 40 bytes (five double fields), and the number of history/prediction elements is $M = 5$. The parameter $r_0$ in the probability of including an object in the broadcasted messages is 100 meters. The message-size control strategy is implemented so that the CBR is kept fixed at $\text{CBR}^* = 0.68\%$. The simulation time is 100 seconds, and all of the results are averaged over this time period across all vehicles in the network.

Figure 7 and Figure 8 show the PER versus distance, when the car density is 25 and 125 vehicles/Km on the 4-Km road topology shown in Figure 2, respectively. The transmission range is assumed to be 500 meters, which is equivalent to the transmission power of 21 dBm. A comparison of the PER values for the cases with and without the control mechanism shows that the message-size control strategy is effective in reducing PER drastically, especially when the car density is high.

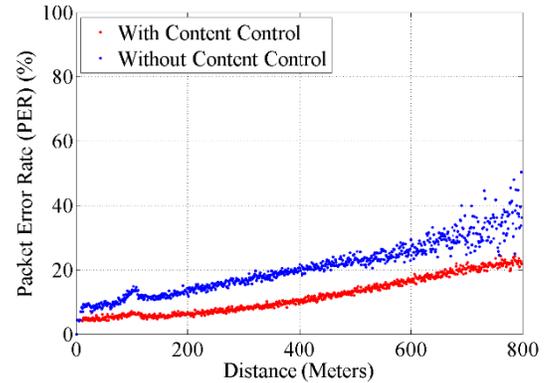

Figure 7: PER versus distance for car density of 25 vehicles per kilometer

As the ultimate goal of the map-exchange protocol is to have a real-time estimate of the positions of the objects at each vehicle, the *information age* (IA) is a measure that could describe how effective this protocol works. The IA is a performance measure that quantifies the *staleness* of the map information at each car. Figure 9 and Figure 10 show the IA versus distance for the network described above, when the car density is 25 and 125 vehicles/Km on the 4-Km road topology shown in Figure 2, respectively. The transmission range is assumed to be 500 meters, which is equivalent to the transmission power of 21 dBm. As these figures show, the information age remains relatively low when the proposed content-control strategy is implemented. It should be noted that the proposed content-control scheme reduces the CBR, which in turn decreases the PER; therefore, more broadcasted messages are delivered to a higher number of vehicles and consequently, the information about more objects in the map is updated more frequently, resulting in the reduction in the IA.

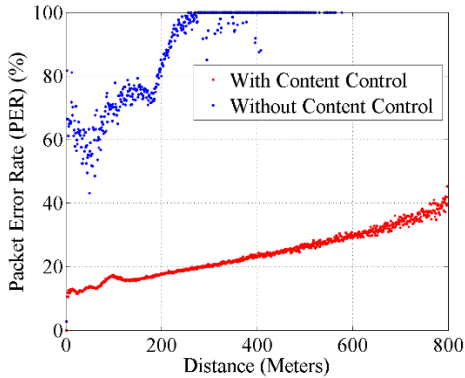

Figure 8: PER versus distance for car density of 125 vehicles per kilometer

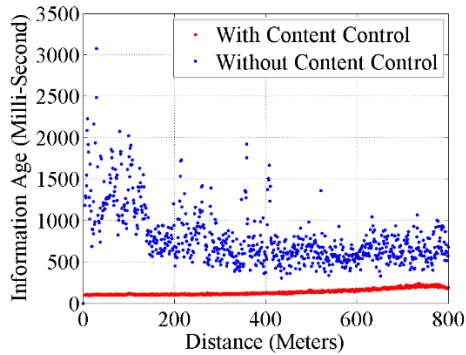

Figure 9: IA versus distance for car density of 25 vehicles per kilometer

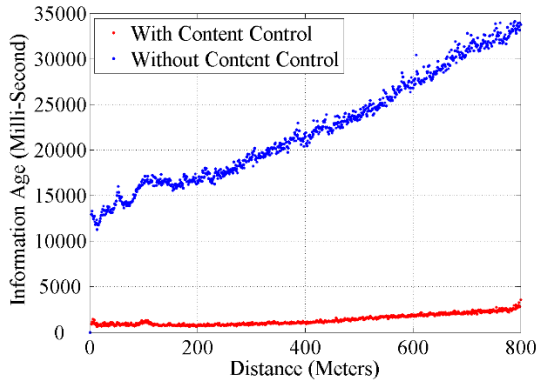

Figure 10: IA versus distance for car density of 125 vehicles per kilometer

## VI. CONCLUSIONS

In this paper, the scalability problem in the conventional communication systems for cooperative automated vehicles was demonstrated through simulation of a typical vehicular network. To resolve this issue, a general communication framework was proposed in which each car creates a multi-resolution map based on the information collected by its local sensing modalities as well as the information received from the maps broadcasted by other vehicles in its communication range. The exchange of map information among cars extends the visibility range of each vehicle beyond the region that can directly be sensed by its sensors. To address the scalability problem of such scheme, a content-control strategy was proposed to adapt the size and content of messages broadcasted by a vehicle based on locally observable network performance measures. The proposed scheme is a complementary technique to existing rate- or range-control mechanisms that are designed for current vehicular-safety networks, but are inadequate for cooperative automated-vehicle cases. The possibility of controlling the content of the messages provides a larger range of adaptation, while still allowing communication among automated vehicles at a reasonable rate. The results of our NS-3 simulations showed that the proposed scheme is scalable and reduces the loss and latency in information delivery compared to other similar schemes.